\documentclass[12pt]{article}
\usepackage{amsmath}
\usepackage{amsfonts}
\usepackage[english]{babel}
\usepackage[ansinew]{inputenc}
\usepackage{graphicx}
\usepackage{graphpap}

\begin{document}

\newcommand{\bm}[1]{\mbox{\boldmath $#1$}}

\newtheorem{defi}{Definition}
\newtheorem{lema}{Lemma}
\newtheorem{proof}{Proof}
\newtheorem{thr}{Theorem}
\newtheorem{corollary}{Corollary}
\newtheorem{proposition}{Proposition}

\def\U{U}
\def\N{\cal N}
\def\tr{\mbox{tr} \,}
\def\Sb{S_{b}}
\def\Db{{\mathfrak D}_{b}}
\def\D{{\mathfrak D}}
\def\SN{\Sigma\cap\mathcal{N}}
\def\gM{g^{(4)}}
\def\id{(\Sigma,g,K)}
\def\idfull{(\Sigma,g,K;\rho,{\bf J})}
\def\mots{S}
\def\KVmotsM{\vec{\Pi}}
\def\MCVmotsM{\vec{H}}
\def\MCmotsM{H}
\def\gmots{\gamma}
\def\MCVmotsS{\vec{p}}
\def\MCmotsS{p}
\def\KVmotsS{\vec{\kappa}}
\def\E{B}
\def\gE{\gamma}
\def\MCVES{\vec{p}_{B}}
\def\MCES{p_{B}}
\def\gl{{g_{0}}}
\def\etal{{\eta_{0}}}
\def\kid{(\Sigma,g,K;N,\vec{Y},\tau)}
\def\R{R}
\def\Rg{{R^{g}}}
\def\C{C}
\def\Rh{{R^{h}}}
\def\KES{\kappa}
\def\hE{\gamma}
\def\MCES{p}

\def\Journal#1#2#3#4{{#1} {\bf #2}, #3 (#4)}
\def\JGP{\em J. Geom. Phys.}
\def\JDG{\em J. Diff. Geom.}
\def\CQG{\em Class. Quantum Grav.}
\def\JPA{\em J. Phys. A: Math. Gen.}
\def\PRD{{\em Phys. Rev.} \bm{D}}
\def\GRG{\em Gen. Rel. Grav.}
\def\IJT{\em Int. J. Theor. Phys.}
\def\PR{\em Phys. Rev.}
\def\RMP{\em Rev. Mod. Phys.}
\def\MNRAS{\em Mon. Not. Roy. Astr. Soc.}
\def\JMP{\em J. Math. Phys.}
\def\DG{\em Diff. Geom.}
\def\CMP{\em Commun. Math. Phys.}
\def\APP{\em Acta Phys. Polon.}
\def\PRL{\em Phys. Rev. Lett.}
\def\ARAA{\em Ann. Rev. Astron. Astroph.}
\def\ANP{\em Annals Phys.}
\def\AP{\em Ap. J.}
\def\APJL{\em Ap. J. Lett.}
\def\MPL{\em Mod. Phys. Lett.}
\def\PREP{\em Phys. Rep.}
\def\AASF{\em Ann. Acad. Sci. Fennicae}
\def\ZP{\em Z. Phys.}
\def\PNAS{\em Proc. Natl. Acad. Sci. USA}
\def\PLMS{\em Proc. London Math. Soth.}
\def\AIHP{\em Ann. Inst. H. Poincar\'e}
\def\ANYAS{\em Ann. N. Y. Acad. Sci.}
\def\SPJ{\em Sov. Phys. JETP}
\def\PAWBS{\em Preuss. Akad. Wiss. Berlin, Sitzber.}
\def\PPLL{\em Phys. Lett. A }
\def\QJRAS{\em Q. Jl. R. Astr. Soc.}
\def\CR{\em C.R. Acad. Sci. (Paris)}
\def\CP{\em Cahiers de Physique}
\def\NC{\em Nuovo Cimento}
\def\AM{\em Ann. Math.}
\def\APP{\em Acta Physica Polonica}
\def\BAMS{\em Bulletin Amer. Math. Soc}
\def\CPAM{\em Commun. Pure Appl. Math.}
\def\PJM{\em Pacific J. Math.}
\def\ATMP{\em Adv. Theor. Math. Phys.}
\def\PRSA{\em Proc. Roy. Soc. A.}
\def\APPT{\em Ann. Poincar\'e Phys. Theory}

\title{On marginally outer trapped surfaces in stationary and
static spacetimes}
\author{Alberto Carrasco$^1$ and Marc Mars$^2$ \\
Facultad de Ciencias, Universidad de Salamanca,\\
 Plaza de la Merced s/n, 37008 Salamanca, Spain \\
$^1$ acf@usal.es, $^2$ marc@usal.es}

\maketitle

\begin{abstract}
In this paper we prove that for any 
spacelike hypersurface 
containing an untrapped barrier
in a stationary spacetime
satisfying the null energy condition,
any marginally outer 
trapped surface cannot lie in the exterior
region where the stationary Killing vector is timelike.
In the static case
we prove that any marginally outer
trapped surface cannot {\it penetrate}
into the exterior
region where the static Killing vector is timelike.
In fact, we prove
these result at an initial data level, without even assuming existence
of a spacetime. The proof relies on a powerful theorem by
Andersson \& Metzger on existence of an outermost marginally outer trapped
surface. 

\end{abstract}

\section{Introduction}\label{introduction}

In 2005 Miao \cite{M} proved a theorem generalizing the
classic uniqueness result of Bunting and Masood-ul-Alam \cite{BM}
for static black holes. This classic theorem proves that a
time-symmetric slice of a static black hole, or more precisely
a three-dimensional asymptotically flat Riemannian manifold
with a totally geodesic boundary where the Killing vector vanishes, must
be a time-symmetric slice of Schwarzschild spacetime.
Miao was able to reach the same conclusion under much weaker assumptions, namely
by replacing the conditions on the boundary from being totally geodesic and 
with vanishing Killing to simply being minimal.
As in Bunting and Masood-ul-Alam's theorem, Miao's
result deals with static, time-symmetric and asymptotically flat
initial data sets for vacuum. 
A key ingredient in Miao's proof was to show that in such an initial
data set, existence of a minimal surface implies the  existence
of a totally geodesic surface where the Killing vector vanishes and moreover,
that the given minimal surface must coincide with it.
Hence, the black hole uniqueness proof implies that the exterior of
the minimal surface must coincide with the exterior
of the time symmetric slice of 
Schwarzschild spacetime outside the black hole. 

Since for static, time-symmetric, vacuum slices the set of points where the 
Killing vector vanishes is known to be a totally geodesic surface, the key
step in Miao's proof can 
be rephrased as saying that
a minimal surface cannot penetrate
the exterior region where the Killing vector is timelike. 

In this paper we extend this result 
in three different directions. Firstly we
allow for non-vanishing matter as long as the null energy condition is satisfied. Secondly,
our initial data sets are no longer required to be time-symmetric. In this case the natural
replacement for minimal surfaces is that of marginally outer trapped surfaces (MOTSs).
And finally we relax the condition of asymptotic flatness to just assuming the 
presence of an asymptotically untrapped barrier (defined below). 
In this general
setting we prove two results, one for the stationary and one for the static case.
In the stationary case, we show that
any bounding MOTS satisfying a suitable
reasonable property cannot lie in 
the outer region where the Killing vector is timelike. The precise
statement is given in Theorem \ref{theorem1}. 
In the static case we can strengthen this result and show that 
no bounding MOTS satisfying the same property
can {\it penetrate} into the exterior region where the Killing 
is timelike. The precise statement is given in Theorem \ref{theorem2}.
These results for MOTSs also hold for weakly outer trapped surfaces.

We emphasize that these results represent an extension of 
Miao's uniqueness theorem in the following sense. In the case
of static, time-symmetric,
asymptotically flat and vacuum initial data sets, an untrapped
barrier always exists and all the conditions
in our theorems are automatically fulfilled. Hence, existence
of a minimal surface implies, from Theorem \ref{theorem1}, that there must exist
a surface of fixed points for the Killing vector. Secondly, the 
non-penetration property (Theorem \ref{theorem2}) 
shows that the given minimal surface
must coincide with this totally geodesic surface and then the
Bunting and Masood-ul-Alam black hole uniqueness proof applies to
show that the region outside the minimal surface must be isometric 
to the time-symmetric slice of Schwarzschild spacetime outside the
black hole.

Theorem \ref{theorem1} 
also generalizes a result 
in \cite{MS} where it was proven
that  strictly stationary spacetimes cannot contain
closed trapped nor marginally (non-minimal) trapped surfaces.
Notice that the definition of
trapped or marginally trapped restricts {\it both}
null expansions of the surface, while weakly outer
trapped surfaces only restrict one of them.

The proof given by Miao relies strongly on the vacuum field equations, so 
we must resort to different methods.
The main technical tool that will allow us to extend  Miao's result in such generality
is a recent powerful theorem
by Andersson and Metzger \cite{AM}, which asserts, roughly speaking, that
the boundary of the weakly outer trapped region in initial data sets is either empty
or a smooth marginally outer trapped surface.
The existence of such an outermost
surface in the minimal case was already known (see \cite{HI} and references
therein) and was in fact an important step in the proof by Miao. With this
generalization
to the non-time-symmetric
setting at hand, it is natural to ask whether Miao's results also extend
and in which sense.

Investigations involving stationary and static spacetimes have followed a general tendency
over the years of relaxing global assumption in time 
and trying to work directly on slabs of spacetimes containing suitable spacelike hypersurfaces.
This is particularly noticeable in black hole uniqueness theorems, where
several conditions can be used to capture the notion of black hole (not all
of them immediately equivalent).
In this paper, we follow this trend and work exclusively
at the initial data level, without even assuming
the existence of a spacetime containing it. In some circumstances
the existence of such spacetime can be proven, for example by using the notion of
Killing development \cite{BC} or by using well-posedness of the Cauchy problem 
and suitable evolution equations for the Killing \cite{Coll}. The former,
however, fails at fixed points (see below for
the definitions) and the second requires specific matter models, not just energy
inequalities as we assume here. Thus, at the level of generality we work on this
paper, the existence of a spacetime cannot always be guaranteed and dealing directly
with initial data sets puts the problem into a more general setting. In particular,
some of our results generalize
known properties of static spacetimes to the initial data setting, which may be of independent
interest.
All the definitions we put forward are therefore stated in terms of initial data
sets. However, since they are motivated by a spacetime point of view, we
often explain briefly the spacetime perspective before giving the definition for the
abstract initial data set.

We finish the introduction with a brief summary of this work.
In Section \ref{preliminaries} we
define initial data set as well as Killing initial data set. Then we
introduce the so-called Killing form, give some of its properties and 
recall the definition of
MOTS in terms of initial data sets. In Section \ref{staticityKID}
we discuss the implications of imposing staticity
on a Killing initial data set and state a number of useful properties of the boundary of the
set where the static Killing vector is timelike, which will be important to prove our main
theorems. Some of the technical work required in this section is
related to the fact that we are not a priori assuming the existence of an spacetime,
and some of the results may be of independent interest.
Finally, Section \ref{mainresults} is devoted to stating and proving the two theorems
discussed above on non-existence of MOTSs in the outer timelike region, one
for the stationary
case and another for the static one.

\section{Preliminaries}
\label{preliminaries}

\subsection{Killing Initial Data (KID)} \label{sKID}

We start with the standard definition of initial data set
(throughout this paper Latin indices run form 1 to 3, Greek indices
run for 0 to 4 and boldface letters denote one-forms).
\begin{defi}
An \textbf{initial data set} $(\Sigma,g,K;\rho,{\bf J})$ is a
3-dimensional connected manifold $\Sigma$
endowed with a Riemannian metric $g$, a symmetric, rank-two
tensor $K$, a scalar $\rho$ and a one-form $\bf{J}$
satisfying
\begin{eqnarray*}
2\rho & = & \Rg+ (\tr  K)^{2}-K_{ij}K^{ij},  \\
-J_{i} & = & D^j(K_{ij}- \tr  K g_{ij}),
\end{eqnarray*}
where $\Rg$ is the scalar curvature and $D$ the covariant derivative of $g$
and $\tr  K= g^{ij}K_{ij}$.
\end{defi}

For simplicity, we will often write $\id$ instead of $\idfull$
when no confusion arises.

In the framework of the Cauchy problem for the Einstein field equations,
$\Sigma$ is an embedded spacelike submanifold of a spacetime
$(M,\gM)$, $g$ is
the induced metric and $K$ is the second fundamental form, which is
defined as
$K(\vec{X},\vec{Y})=- \bm{n} ( \nabla_{\vec{X}}\vec{Y} )$
where $\nabla$ is the covariant derivative of $\gM$,
$\bm{n}$ is a unit future directed normal one-form to $\Sigma$ and
$\vec{X},\vec{Y}$ are
arbitrary vector fields tangent to $\Sigma$, i.e.
$\vec{X},\vec{Y} \in \mathfrak{X}(\Sigma)$.
Let $G_{\mu\nu}$ be the Einstein tensor of $\gM$.
The initial data energy density $\rho$ and energy flux ${\bf J}$
are defined by $\rho \equiv G_{\mu\nu}n^{\mu}n^{\nu}, J_{i} \equiv
-G_{\mu\nu}n^{\mu}e_{i}^{\nu}$, where $\{ \vec{e}_i \}$ is a basis vector
field for $\mathfrak{X}(\Sigma)$.
When $\rho=0$ and ${\bf J}=0$, the initial data set
is said to be {\bf  vacuum}.

As remarked in the Introduction
we will regard initial data sets as abstract objects
on their own, independently of the existence of a spacetime,
unless explicitly stated.

Consider now a spacetime
$(M,\gM)$ admitting a local isometry generated by
a Killing vector field $\vec{\xi}$, i.e.
$\mathcal{L}_{\vec{\xi} \, }g^{(4)}_{\mu\nu} = \nabla_{\mu}\xi_{\nu} +
\nabla_{\nu}\xi_{\mu} = 0$,
where  $\mathcal{L}_{\vec{v}}$ is the Lie derivative along
$\vec{v}$ and let $\id$ be an initial data set in this spacetime.
We can decompose $\vec{\xi}$ along $\Sigma$ into a normal and
a tangential component as
\begin{equation}\label{killingdecomposition}
\vec{\xi}=N\vec{n}+Y^{i}\vec{e}_{i},
\end{equation}
where $N = -\xi^{\mu}n_{\mu}$.
Inserting this into the Killing equations and performing a 3+1 splitting
on $\id$ it follows (see \cite{Coll}, \cite{BC}),
\begin{eqnarray}
2NK_{ij} +  2D_{(i}Y_{j)}=0, \hspace{98mm} \label{kid1} \\
\mathcal{L}_{\vec{Y}}K_{ij}  +  D_{i}D_{j}N=N\left(
\Rg_{ij}+ \tr K K_{ij}-2K_{il}K_{j}^{l} \right) - N\left(
\tau_{ij}-\frac{1}{2}g_{ij}(\tr \tau-\rho) \right), \label{kid2}
\end{eqnarray}
where $\Rg_{ij}$ is the Ricci tensor of $g$,
$\tau_{ij} \equiv G_{\mu\nu}e_{i}^{\mu}e_{j}^{\nu}$ are the
remaining components of the Einstein tensor and
$\tr \tau=g^{ij}\tau_{ij}$. Thus, the following definition
of Killing initial data becomes natural \cite{BC}.

\begin{defi} An initial data set $\idfull$ endowed with a scalar $N$, a
vector $\vec{Y}$ and a symmetric tensor
$\tau_{ij}$ satisfying equations
(\ref{kid1}) and (\ref{kid2}) is called a \textbf{Killing initial data}
\textit{(KID)}.
\end{defi}
In particular, if a KID has $\rho=0$, ${\bf J}=0$ and $\tau=0$ then it is said to be a
{\bf vacuum KID}.

A point $p \in \Sigma$  where $N=0$ and $\vec{Y}=0$ is called a {\bf fixed point}.
This name is motivated by the fact that when the KID is embedded into a spacetime with a
local isometry, the corresponding Killing vector $\vec{\xi}$ vanishes at $p$ and the isometry has
a fixed point there.

A natural question regarding KIDs is whether they can be embedded into a spacetime
$(M,\gM)$ such that $N$ and $\vec{Y}$ correspond to a Killing vector $\vec{\xi}$.
The simplest case
where existence is guaranteed involves ``transversal'' KIDs, i.e. when
$N\ne0$ everywhere. Then, the following spacetime, called
\textit{Killing development} of $(\Sigma,g,K)$,
 can be constructed
\begin{equation}
\label{killingdevelopment}
\left( \Sigma\times\mathbb{R},\quad g^{(4)}=-\hat{\lambda}dt^{2}
+2\hat{Y}_{i}dtdx^{i}+ \hat{g}_{ij}dx^{i}dx^{j} \right)
\end{equation}
where
\begin{eqnarray}\label{killingdevelopment2}
\hat{\lambda}(t,x^{i})\equiv (N^2 - Y^i Y_i )(x^{i}), & \hat{g}_{ij}(t,x^{k})\equiv
g_{ij}(x^{k}), & \hat{Y}^{i}(t,x^{j})\equiv Y^{i}(x^{j}).
\end{eqnarray}
Notice that $\partial_{t}$
is a complete Killing field with orbits diffeomorphic to $\mathbb{R}$
which, when evaluated on $\Sigma \equiv \{t=0 \}$
decomposes as $\partial_{t}=N\vec{n}+Y^{i}\vec{e}_{i}$, in agreement with
(\ref{killingdecomposition}). The Killing development is the unique
spacetime with these properties.
Further details can be found in \cite{BC}. Notice also that the Killing development
can be constructed for any connected subset of $\Sigma$ where $N \neq 0$ everywhere.

\subsection{Killing Form on a KID}\label{subsectionkillingform}

A useful object in spacetimes with a Killing vector $\vec{\xi}$ is
the two-form $\nabla_{\mu}\xi_{\nu}$, usually called {\bf Killing form} or also
Papapetrou field. This tensor will play a relevant role  below. Since
we intend to work directly on the initial data set,
we need to define a suitable tensor on $\id$
which corresponds to the Killing form  whenever a spacetime is present.
Let $\kid$ be a KID in $(M,\gM)$. Clearly we need
to restrict and decompose  $\nabla_{\mu}\xi_{\nu}$ onto $\kid$
and try to get an expression in terms of $N$ and $\vec{Y}$ and its spatial
derivatives. In order to use
(\ref{killingdecomposition}) we first extend
$\vec{n}$ to a neighbourhood of $\Sigma$ as a unit and hypersurface
orthogonal, but otherwise arbitrary, vector field
(the final expression we obtain will be independent of this
extension), and define $N$ and $\vec{Y}$ so that $\vec{Y}$ is
orthogonal to $\vec{n}$ and (\ref{killingdecomposition}) holds.
Taking covariant derivatives we find
\begin{equation}
\label{KF}
\nabla_{\mu}\xi_{\nu}=\nabla_{\mu}Nn_{\nu}+N\nabla_{\mu}n_{\nu}+\nabla_{\mu}Y_{\nu}.
\end{equation}
Notice that, by construction,
$\nabla_{\mu}n_{\nu} |_{\Sigma}=K_{\mu\nu}-n_{\mu}a_{\nu} |_{\Sigma}$
where $a_{\nu}=n^{\alpha}\nabla_{\alpha}n_{\nu}$ is the acceleration of $\vec{n}$.
To elaborate $\nabla_{\mu} Y_{\nu}$ we recall that $D$-covariant derivatives
correspond to spacetime covariant derivatives projected
onto $\Sigma$. Thus, it follows easily
\begin{eqnarray*}
\nabla_{\mu}Y_{\nu} |_{\Sigma} =
D_{\mu}Y_{\nu}-n_{\mu}\left( n^{\alpha}\nabla_{\alpha}Y_{\beta} \right)
h^{\beta}_{\nu}+
K_{\mu\alpha}Y^{\alpha}n_{\nu}+n_{\mu}n_{\nu}n^{\alpha}n^{\beta}\nabla_{\alpha}Y_{\beta} |_{\Sigma},
\end{eqnarray*}
where $h^{\mu}_{\nu}=\delta^{\mu}_{\nu}+n^{\mu}n_{\nu}$
is the
projector orthogonal to $\vec{n}$,
and $D_{\mu} Y_{\nu} \equiv h^{\alpha}_{\mu} h^{\beta}_{\nu} \nabla_{\alpha} Y_{\beta}$.
Substitution into (\ref{KF}) gives

\begin{eqnarray}
\label{killingform1}
\nabla_{\mu}\xi_{\nu}\big|_{\Sigma}&=&n_{\nu}\left( D_{\mu}N+K_{\mu\alpha}Y^{\alpha} \right) -
n_{\mu}\left( Na_{\nu}+n^{\alpha}h^{\beta}_{\nu} \nabla_{\alpha}Y_{\beta}
\right) + \nonumber\\
&&\quad +
(D_{\mu}Y_{\nu}+NK_{\mu\nu}) + n_{\mu}n_{\nu}\left(
n^{\alpha}n^{\beta}\nabla_{\alpha}Y_{\beta}-
n^{\alpha}\nabla_{\alpha}N \right) |_{\Sigma}.
\end{eqnarray}
The Killing equations then require $
n^{\alpha}n^{\beta}\nabla_{\alpha}Y_{\beta} |_{\Sigma}=
n^{\alpha}\nabla_{\alpha}N |_{\Sigma} $ and
$ D_{\mu}N+K_{\mu\alpha}Y^{\alpha}|_{\Sigma}= Na_{\mu}+n^{\alpha}h^{\beta}_{\mu} \nabla_{\alpha}Y_{\beta} |_{\Sigma}$,
so that (\ref{killingform1}) becomes, after using  (\ref{kid1}),
\begin{equation}
\label{killingform2}
\left. \nabla_{\mu}\xi_{\nu} \right |_{\Sigma} =
\left . n_{\nu}\left( D_{\mu}N+K_{\mu\alpha}Y^{\alpha} \right) -
n_{\mu}\left( D_{\nu}N+K_{\nu\alpha}Y^{\alpha} \right) +
\frac12\left( D_{\mu}Y_{\nu}-D_{\nu}Y_{\mu} \right) \right |_{\Sigma}.
\end{equation}
This expression involves solely objects defined on $\Sigma$. However,
it still involves four-di\-men\-sio\-nal objects. In order to work
directly on the KID, we introduce an auxiliary
four-dimensional vector space on each point of $\Sigma$ as follows (we
stress that we are {\it not} constructing a spacetime, only a
Lorentzian vector space attached to each point on the KID).

At every $p\in\Sigma$ define the vector space $V_p=T_{p}\Sigma\times\mathbb{R}$,
and endow this space with the Lorentzian metric
$\gl |_p =g |_p\oplus\left( -\delta \right)$, where ${\delta}$ is the
canonical metric on $\mathbb{R}$. Let $\vec{n}$ be the unit
vector tangent to the fiber $\mathbb{R}$.
 Having a metric we
can lower and raise indices of tensors in $T_{p}\Sigma\times\mathbb{R}$.
In particular define ${\bf n}=\gl(\vec{n},\cdot)$. Covariant tensors
$Q$ on $T_p \Sigma$ can be canonically extended to tensors of the same rank on
$V_p = T_{p}\Sigma\times\mathbb{R}$ (still denoted with the same symbol) simply by noticing
that any vector in $V_p$ is
of the form $\vec{X} + a \vec{n}$, where $\vec{X} \in T_p \Sigma$ and $a \in \mathbb{R}$. The extension
is defined (for a rank $m$ tensor) by $Q (\vec{X_1} + a_1 \vec{n}, \cdots, \vec{X}_m + a_m \vec{n})
\equiv Q(\vec{X_1}, \cdots, \vec{X}_m)$. In index notation, this extension will
be expressed simply by changing Latin to Greek indices. It is clear that
the collection of $\left( T_{p}\Sigma\times\mathbb{R},\gl \right)$ at
every $ p\in\Sigma$ contains no more information than just $(\Sigma,g)$.

Motivated by (\ref{killingform2}), we can define Killing form
directly in terms of objects on the KID
\begin{defi}
The {\bf Killing form on a KID} is the 2-form $F_{\mu\nu}$
defined on $\left( T_{p}\Sigma\times\mathbb{R}, \gl \right)$ introduced above
given by
\begin{equation}\label{killingform}
F_{\mu\nu}= n_{\nu}\left( D_{\mu}N+K_{\mu\alpha}Y^{\alpha} \right) -
n_{\mu}\left( D_{\nu}N+K_{\nu\alpha}Y^{\alpha} \right) +
f_{\mu\nu},
\end{equation}
where $f_{\mu\nu}=D_{[\mu}Y_{\nu]}$ and brackets denote antisymmetrization.
\end{defi}
In a spacetime setting it is well-known that for a non-trivial Killing vector
$\vec{\xi}$, the Killing form cannot vanish on a fixed point. Let us show that
the same happens in the KID setting.
\begin{lema}
\label{FixedPointF=0}
Let $\kid$ be a KID and $p \in \Sigma$ a fixed point, i.e. $N |_p = 0$ and
$\vec{Y} |_p =0$. If $F_{\mu\nu} |_p = 0$ then $N$ and $\vec{Y}$ vanish identically on
$\Sigma$.
\end{lema}
{\it Proof.}
The aim is to obtain a suitable system of equations and show that, under the
circumstances of the Lemma,  the solution must be identically zero.
Decomposing $D_{i}Y_{j}$ in symmetric
and antisymmetric parts,
\begin{equation}\label{DY}
D_{i}Y_{j}=-NK_{ij}+f_{ij},
\end{equation}
and inserting into (\ref{kid2}) gives
\begin{equation}\label{DDN}
D_{i}D_{ j}N=NQ_{i j}-Y^{l}D_{l}K_{i j}-K_{il}f_{ j}^{l}-K_{ jl}
f_{i}^{l},
\end{equation}
where $Q_{i j}=\Rg_{i j}+KK_{i j}-\tau_{i j} + \frac{1}{2}g_{i j}(\tr \tau-\rho)$.
In order to find an equation for $D_{l}f_{i j}$, we take a derivative of
(\ref{kid1}) and write the three equations obtained by cyclic
permutation. Adding two of them and substracting the third one, we
find, after using the Ricci and first Bianchi identities,
$D_l D_i Y_j = \Rg_{klij} Y^k +
D_{j} ( N K_{li}) - D_i (N K_{lj}) - D_l (N K_{ij}) 
$. Taking the antisymmetric part,
\begin{equation}\label{DF}
D_{l} f_{i j}=\Rg_{k l i j}Y^{k}+D_{ j}NK_{li}-D_{i}NK_{l j} +
ND_{ j}K_{li}-ND_{i}K_{l j}.
\end{equation}
If $F_{\mu\nu} |_p=0$, it follows that $f_{ij} |_p =0$ and $D_i N |_p =0$.
The equations given by (\ref{DY}), (\ref{DDN}) and (\ref{DF}) is a system of
PDEs for the unknowns $N$, $Y_i$ and $f_{ij}$ written in normal form. It follows
(see e.g. \cite{Eisenhart}) that the vanishing of $N$, $D_iN$, $Y_i$ and $f_{ij}$
at one point implies its vanishing everywhere (recall that $\Sigma$ is connected). $\hfill \square$

\subsection{Canonical Form of Null two-forms}
\label{canonical}

Let $F_{\mu\nu}$ be an arbitrary two-form on a spacetime $(M,g^{(4)})$.
It is well-known that the only two non-trivial scalars that can be constructed from
$F_{\mu\nu}$ are $I_{1}=F_{\mu\nu}F^{\mu\nu}$ and
$I_{2}= F^{\star}_{\mu\nu}F^{\mu\nu}$,
where $F^{\star}$ is the Hodge dual of $F$, defined by
$F^{\star}_{\mu\nu}=\frac12\eta_{\mu\nu\alpha\beta}F^{\alpha\beta}$,
with $\eta_{\mu\nu\alpha\beta}$ being the volume form of $(M,g^{(4)})$.
When both scalars vanish, the two-form is called \textit{null}.
Later on, we will encounter Killing forms which are null and we will exploit
the following well-known algebraic decomposition
which gives its {\bf canonical form}, see e.g. \cite{Israel} for a proof.
\begin{lema}
\label{lemmacanonicalform}
A null two-form $F_{\mu\nu}$ at a point $p$ can be decomposed as
\begin{equation}\label{canonicalform}
F_{\mu\nu} |_{p} =l_{\mu} w_{\nu}-l_{\nu}w_{\mu} |_{p},
\end{equation}
where $\vec{l} |_p$ is null vector satisfying
$F_{\mu\nu}l^{\mu} |_p =0$ and  $\vec{w} |_p$ is spacelike and orthogonal to
$\vec{l} |_p$.
\end{lema}

\subsection{Marginally Outer Trapped Surfaces (MOTSs)}\label{sMOTS}
Let $\mots$ be a smooth orientable,
codimension 2, embedded submanifold of $(M,\gM)$
with positive definite first fundamental form $\gmots$.
Let $\KVmotsM$ denote the second
fundamental form-vector of $\mots$ as a submanifold of $M$, defined as
$\KVmotsM(\vec{X},\vec{Y}) =-(\nabla_{\vec{X}}\vec{Y} )^{\bot}$,
$\forall \vec{X},\vec{Y}\in \mathfrak{X}(S)$ and define
the mean curvature vector of $\mots$ in $M$ as
$\MCVmotsM={\gmots}^{AB}\KVmotsM_{AB}$ $(A,B,C...=2,3)$.

The normal bundle of $\mots$
admits a basis $\{\vec{l}, \vec{k} \}$ of smooth, null and future directed
vectors partially normalized to satisfy $l^{\mu}k_{\mu}=-2$. The mean
curvature vector decomposes as
 $\MCVmotsM=-\frac{1}{2} ( \theta_{\vec{k} \,}\vec{l}+
\theta_{\vec{l} \, }\vec{k} )$, where
$\theta_{\vec{l}}\equiv \MCmotsM^{\mu}l_{\mu}$
and $\theta_{\vec{k}}\equiv \MCmotsM^{\mu}k_{\mu}$ are the null expansions of $\mots$.

\begin{defi}\label{mots1} A closed (i.e. compact and without boundary) surface
$\mots$ in a spacetime $(M,\gM)$
is a {\bf marginally outer trapped surface}
\textit{(MOTS)} if $\MCVmotsM$ is proportional to one of the elements of the null basis of its
normal bundle.
\end{defi}
{\em Remark.} The null vector to which $\MCVmotsM$ is proportional is called
$\vec{l}$ and it points to
what is called the \textit{outer} direction. In other words a surface $\mots$
 is a MOTS iff $\theta_
{\vec{l}}=0$. Note that the term outer does not refer to a direction singled out a priori.

According to the philosophy of this work, we need a definition of MOTS
in terms of initial data. Let $\id$ be an initial data set for $(M,\gM)$
and $\mots \subset \Sigma$ an oriented, embedded codimension one submanifold
of $\Sigma$. Such an object is simply called ``surface'' throughout this work.
If the initial data set lies in a spacetime, let
$\KVmotsS_{AB}$ be
the  second fundamental form-vector of $\mots$ as a
a submanifold of $\Sigma$, and $\MCVmotsS=\gmots^{AB} \KVmotsS_{AB}$
the mean curvature vector. A standard formula relates the spacetime
mean curvature to these objects by
\begin{equation}\label{h}
\MCVmotsM=\MCVmotsS - \gmots^{AB} K_{AB}\vec{n}.
\end{equation}
where $K_{AB}$ is the pull-back of $K_{ij}$ onto $\mots$. Let $\vec{m}$
be the unique (up to orientation) unit normal to $\mots$
tangent to $\Sigma$. Then, a suitable choice for null basis  $\{ \vec{l}, \vec{k} \}$
is $\vec{l}=\vec{n}+\vec{m}$ and $\vec{k}=\vec{n}-\vec{m}$.
Multiplying (\ref{h}) by $\vec{l}$ we find
$\theta_{\vec{l}}= \MCmotsS+\gmots^{AB}K_{AB}$, after writing
$\MCVmotsS = \MCmotsS \vec{m}$. All objects are now intrinsic to $\Sigma$,
which leads to the following standard definition.
\begin{defi}
\label{defiMOTS}
A closed surface $S$ in an initial data set $\id$ is a \textbf{
marginally outer trapped surface} \textit{(MOTS)} iff
\begin{equation}
\label{motsSigma}
\MCmotsS+\gmots^{AB}K_{AB}=0,
\end{equation}
where $\gmots$ is the induced metric on $\mots$, $K_{AB}$ is the pull-back of $K_{ij}$ to $\mots$
and $\MCmotsS$ is the mean curvature
of $\mots$ w.r.t. a unit normal direction $\vec{m}$,
called the outer direction.
\end{defi}

For MOTSs, equation (\ref{motsSigma}) singles out which one is
the outer direction (when $p= \gmots^{AB} K_{AB} =0$ both directions
are outer according to this definition). If for some reason one can
single out an outer direction for a given  surface $\mots$, then we shall say that
$\mots\subset\Sigma$ is {\bf weakly outer trapped}
iff $p+\gmots^{AB}K_{AB}\le0$, where $p$ is the mean curvature of
$\mots$ in $\Sigma$ w.r.t. the outer direction.

In this work we shall be concerned with a particular class of MOTS having the property
of being boundaries of domains. To be more precise we first need to concept of barrier surface.
\begin{defi} \label{barrier}
Let $\id$ be an initial data set. A closed surface $\Sb\subset \Sigma$  is called  an
{\bf untrapped barrier surface} provided $\Sb$ is the boundary of
an open domain $\Db$ and  $\MCmotsS+\gmots^{AB}K_{AB} |_{\Sb} >0$,  where
the unit normal defining $\MCmotsS$ points outwards of
$\Db$
\end{defi}

We can now restrict the class of MOTS considered in the paper.
\begin{defi}\label{boundingMOTS}
Let $\id$ be an initial data set with an untrapped barrier surface $\Sb = \partial \Db$.
A closed surface $\mots\subset \Db$
is called a \textbf{bounding MOTS} iff
it is the boundary of an open domain $\D \subset \Db$ and
$\MCmotsS+\gmots^{AB}K_{AB}\big|_{S}=0$ with $\MCmotsS$ being the mean curvature w.r.t. the normal vector
to $\mots$ pointing outwards of $\D$.
\end{defi}
{\em Remark.}  A surface $\mots=\partial\D$  satisfying
$\MCmotsS+\gmots^{AB}K_{AB}\big|_{S}\le0$,
is called a {\bf bounding weakly outer trapped surface}.

\section{Staticity of a KID}
\label{staticityKID}

Most of the results in this paper involve Killing initial data having a static Killing vector.
The concept of staticity is a spacetime one. As usual, we will rewrite the
staticity conditions directly in terms of the initial data set and then will put
forward a definition of static KID.

\subsection{Static KID}

Recall that a spacetime is stationary if it admits
a Killing field $\vec{\xi}$ which is timelike in some
non-empty set. If furthermore, $\vec{\xi}$ is integrable,
i.e.
\begin{equation}
\label{integrability-1}
{\bm \xi}\wedge d{\bm \xi}=0
\end{equation}
the spacetime is called {\it static}. Static spacetimes
can be locally foliated by hypersurfaces orthogonal to $\vec{\xi}$.

Let us now decompose (\ref{integrability-1}) according to (\ref{killingdecomposition}).
By taking the normal-tangent-tangent part
(to $\Sigma$) and the completely tangential part (the other components
are identically zero by antisymmetry) we find
\begin{equation}\label{static5}
ND_{[i}Y_{j]}+2Y_{[i}D_{j]}N+2Y_{[i}K_{j]l}Y^{l}=0,
\end{equation}
\begin{equation}\label{statictwo}
Y_{[i}D_{j}Y_{k]}=0.
\end{equation}
Since this objects involve only objects on the KID, the following definition becomes
natural.
\begin{defi}
A KID $\kid$ satisfying  (\ref{static5}) and (\ref{statictwo})
is called an {\bf integrable KID}.
\end{defi}

Equations (\ref{static5}) and (\ref{statictwo}) together with equation (\ref{kid1}) yield the
following useful relation, valid everywhere on $\Sigma$,
\begin{equation}\label{static1}
\lambda D_{[i}Y_{j]}+Y_{[i}D_{j]}\lambda=0,
\end{equation}
where $\lambda = N^2 - Y^2$. If a spacetime
containing the KID exists, $\lambda$ is precisely minus the squared norm of the
Killing vector, $\lambda = - \xi^{\alpha} \xi_{\alpha}$. Therefore, if
$\lambda>0$ in some non-empty set of the KID, the Killing vector is
timelike in some
non-empty set of the spacetime. Hence
\begin{defi}
A {\bf static KID} is an integrable KID with $\lambda>0$ in some non-empty set.
\end{defi}

\subsection{Killing Form of a Static KID}\label{subsectionFonstaticKID}

In Subsection \ref{canonical} we introduced the invariant scalars $I_1$ and $I_2$
for any two-form in a spacetime. In this section we find their explicit
expressions for the Killing form of an integrable KID in the region $\{\lambda> 0 \}$.

Although non-necessary, we will pass to the Killing development since this simplifies
the proofs. We start with a lemma concerning the integrability of the Killing vector
in the Killing development.
\begin{lema}
\label{integrability}
The Killing vector field associated with the Killing development of an integrable KID is also integrable.
\end{lema}
{\em Proof.} Let $\kid$ be an integrable KID. Suppose the Killing development
(\ref{killingdevelopment}) of a suitable open
set of $\Sigma$. Using $\vec{\xi} = \partial_t$ it follows
\begin{equation}\label{xidxi}
{\bm \xi}\wedge d{\bm \xi}=-\hat{\lambda}\partial_{i}\hat{Y}_{j}dt\wedge dx^{i}\wedge dx^{j} -
\hat{Y}_{i}\partial_{j}\hat{\lambda}dt\wedge dx^{i}\wedge dx^{j} +
\hat{Y}_{i}\partial_{j}\hat{Y}_{k}dx^{i}\wedge dx^{j}\wedge dx^{k},
\end{equation}
where $\hat{\lambda}$, $\hat{{\bf Y}}$ and $\hat{g}$ are defined in (\ref{killingdevelopment2}).
Integrability of $\vec{\xi}$ follows directly from
(\ref{statictwo}) and (\ref{static1}). $\hfill \square$\\

The following lemma gives the explicit expressions for $I_{1}$ and $I_{2}$.
\begin{lema}
\label{lemmaI1}
The invariants of the Killing form in a static KID in the region
$\{ \lambda >0 \}$
read
\begin{equation}\label{I1}
I_{1}=-\frac{1}{2\lambda}\left( g^{ij}-\frac{Y^{i}Y^{j}}{N^{2}}
\right)D_{i}\lambda D_{j}\lambda,
\end{equation}
and
\begin{equation}\label{I2}
I_{2}=0.
\end{equation}
\end{lema}

{\em Remark.} By continuity $I_{2}\big|_{\partial\{\lambda>0\}}=0$.

{\em Proof.} Suppose a static KID $\kid$. In $\{ \lambda>0 \}$
we have necessarily $N\neq 0$, so we can
construct the Killing development of this set, $(\{ \lambda>0 \},g^{(4)},K)$ and
introduce the so-called Ernst one-form, as
$\sigma_{\mu}=\nabla_{\mu} \lambda -i\omega_{\mu}$
where $\omega_{\mu}=\eta^{(4)}_{\mu\nu\alpha\beta}\xi^{\nu}\nabla^{\alpha}\xi^{\beta}$
is the twist of the Killing field
($\eta^{(4)}$ is the volume form of the Killing development).
The Ernst one-form satisfies the identity (see e.g. \cite{Mars})
$\sigma^{\mu}\sigma_{\mu}=-\lambda\left( F_{\mu\nu}+iF^{\star}_{\mu\nu} \right)
\left( F^{\mu\nu}+i{F^{\star}}^{\mu\nu} \right)$, which in the static case (i.e.
$\omega_{\mu}=0$) becomes
$\nabla_{\mu} \lambda \nabla^{\mu} \lambda=-2\lambda\left(F_{\mu\nu}F^{\mu\nu}
+iF_{\mu\nu}{F^{\star}}^{\mu\nu } \right)$
where the identity $F_{\mu\nu}F^{\mu\nu}=-F^{\star}_{\mu\nu}{F^{\star}}^{\mu\nu}$ has been used.
The imaginary part immediately gives (\ref{I2}). The real part gives
$I_{1}=-\frac{1}{2\lambda}|\nabla \lambda|^{2}$.
Taking coordinates $\{t,x ^{i}\}$ adapted to the Killing field $\partial_{t}$, it follows from (\ref{killingdevelopment2}) that
$|\nabla\lambda|^{2}={g^{(4)}}^{ij}\partial_{i}\lambda\partial_{j}\lambda$.
It is well-known (and easily checked) that the contravariant spatial components of $g^{(4)}$
are ${g^{(4)}}^{ij} = g^{ij}-\frac{Y^{i}Y^{j}}{N^2}$, where $g^{ij}$ is the inverse of $g_{ij}$
and (\ref{I1}) follows. $\hfill \Box$

This Lemma allows us to prove the following result on the value of $I_1$
on the fixed points at the closure of $\{ \lambda > 0 \}$. Notice that $\partial \{\lambda>0 \} \subset
\overline{\{N \neq 0\}}$. Since the result involves points where $N$ vanishes, we cannot rely
on the Killing development for its proof, and an argument directly on the initial data set
is needed.

\begin{lema}
\label{I1<0}
Let $p \in \overline{\{\lambda >0 \}}$ be a fixed point of a static KID, then $I_1 |_p < 0$.
\end{lema}

{\it Proof.} We first show that $I_1  \leq 0$ on $\{ \lambda >0 \}$, which implies
that $I_1 |_p \leq 0$ by continuity.
Let $ q\in \{ \lambda>0  \}\subset\Sigma$
and define the vector $\vec{\xi} \equiv N\vec{n}+\vec{Y}$ on the
vector space $(V_q, \gl)$ introduced above.
Since $\vec{\xi}$ is timelike at $q$, we can
introduce its orthogonal projector
$h_{\mu\nu}=\gl_{\mu\nu}+\frac{\xi_{\mu}\xi_{\nu}}{\lambda}$ which is obviously
positive semi-definite. If we pull it back
onto $T_q \Sigma$ we obtain the positive definite orbit space metric
\begin{equation}\label{quotientmetric}
h_{ij}=g_{ij}+\frac{Y_{i}Y_{j}}{\lambda},
\end{equation}
whose inverse corresponds precisely to
the term in brackets in (\ref{I1}) and $I_{1} |_q  \leq 0$ follows.

It only remains to show that $I_1 |_p$ cannot be zero. We argue by contradiction.
Assuming  that  $I_1 |_p =0$ and using $I_2 |_p =0$ by Lemma \ref{lemmaI1}, it follows
that $F_{\mu\nu}$ is null at $p$. Lemma \ref{lemmacanonicalform} implies the
existence of a null vector $\vec{l}$ and a spacelike vector $\vec{w}$ on $V_p$ such
that (\ref{canonicalform}) holds. Since $\vec{w}$ is defined up to
an arbitrary additive vector proportional to $\vec{l}$, we can choose
$\vec{w}$ normal to $\vec{n}$ without loss of generality. Decompose
$\vec{l}$ as $\vec{l}=a\left( \vec{x} + \vec{n} \right)$
with $x^{\mu}x_{\mu}=1$. We know from Lemma \ref{FixedPointF=0}
that $a \neq 0$ (otherwise $F_{\mu\nu} |_p =0$ and $\{ \lambda > 0 \}$ would be empty).
Expression (\ref{killingform}) and the canonical form (\ref{canonicalform}) yield
\begin{equation*}
F_{\mu\nu} |_p  = 2n_{ [ \nu}D_{\mu ] }N + D_{ [ \mu}Y_{\nu ] } |_p = 2a \left( x_{ [ \mu}
w_{\nu ]}+n_{ [ \mu}w_{\nu ] } \right).
\end{equation*}
The purely tangential and normal-tangential components of this equation give, respectively
\begin{eqnarray}
  D_{i}Y_{j}=2a x_{[i}w_{j]}, \quad D_{i}N=aw_{i}, \label{ab}
\end{eqnarray}
where $w_{i}$ is the projection of $w_{\mu}$ to $T_p \Sigma$.
These equations yield the contradiction.
Indeed, take $\vec{v}$ be a geodesic vector field in $\Sigma$,
non-zero at $p$. From $\lambda = N^2 - Y^2$
and the fact that $p$ is a fixed point, (\ref{ab}) easily implies
\begin{equation*}
v^{i}D_{i} \left( v^{j}D_{j} \lambda \right)\big|_{p} = -2a^{2} [
w^{i}w_{i}\left( v^{j}x_{j} \right)^{2} -
2x^{i}w_{i}v^{j}w_{j}v^{k}x_{k} ] =
-2 a^2 w^i w_i \left ( v^j v_j \right )^2 < 0,
\end{equation*}
where, in the second equality we
we used $x^{i}w_{i}=0$, which follows from $\vec{w}$ being
orthogonal to $\vec{l}$.
Being $\vec{v}$  arbitrary (non-zero),
it follows that $\lambda$ has a maximum at $p$, where it vanishes. But this
contradicts
the fact that $p\in\partial\{ \lambda>0 \}$, so that there are points infinitesimally near $p$
with positive $\lambda$. $\hfill \Box$

\subsection{Properties of $\partial \{ \lambda>0 \}$ on a Static KID}

In this subsection we will show that, under suitable conditions, the boundary
of the region $\{\lambda> 0\}$ is a smooth surface. Let us first of all
recall an interesting Lemma
concerning Killing horizons in spacetimes $(M,\gM)$ with an integrable
Killing field $\vec{\xi}$. Recall that a Killing horizon is
a null hypersurface $\mathcal{N}_{\vec{\xi}}$ of $M$ such that the local
isometry generated by $\vec{\xi}$ acts freely on $\mathcal{N}_{\vec{\xi}}$
(i.e. such that the hypersurface is invariant but not pointwise invariant anywhere)
and such that $\vec{\xi}$ is null on $\mathcal{N}_{\vec{\xi}}$.
The Vishveshwara-Carter Lemma reads (see \cite{C} for this form of the statement and its proof).
\begin{lema}[Vishveshwara-Carter \cite{VC}, 1968-69]\label{carter}
Let $(M,\gM)$ be a spacetime with an integrable Killing vector field
$\vec{\xi}$. Then, the set $\mathcal{N}_{\vec{\xi}} \equiv
\partial \{ \lambda>0 \}\cap \{ \vec{\xi}\ne0 \}$, if non-empty, is a Killing horizon.
\end{lema}
We now state our first result on the smoothness of $\partial \{ \lambda > 0 \}$.

\begin{lema}\label{surfaceNoFixed}
Let $\kid$ be a static KID
and assume that the set $\E=\partial\{ \lambda>0 \} \cap \{ N \neq 0 \}$
is non-empty. Then $\E$ is a smooth surface.
\end{lema}
{\it Proof.}
Since $N |_{\E}\ne0$,
we can construct the Killing development (\ref{killingdevelopment})
of a suitable neighbourhood of $\E\subset\Sigma$ satisfying $N \neq 0$ everywhere.
Moreover, by Lemma \ref{integrability} $\vec{\xi} = \partial_t$ is integrable.
Applying the Vishveshwara-Carter Lemma, it follows
that $\mathcal{N}_{\vec{\xi}}$
is a null hypersurface and therefore transverse to $\Sigma$, which is spacelike.
Thus, $\E = \Sigma \cap  \mathcal{N}_{\vec{\xi}}$
is a smooth surface. $\hfill \Box$

This Lemma states that the boundary of $\{ \lambda > 0 \}$ is smooth on the set of
non-fixed points. In fact, for the case of boundaries
having at least one fixed point, an explicit defining function for this surface
on the subset of non-fixed points can be given. This will be useful later.

\begin{proposition}\label{maldito}
Let $\kid$ be a static KID. If
a connected component of $\partial\{\lambda>0 \}$ contains at least one
fixed point, then $D_{i}\lambda\neq 0$ on all non-fixed
points in that connected component.
\end{proposition}

{\em Proof.} Let $U$ be the set of non-fixed points in one of the connected
components under consideration. This set is obviously open.
Constructing the Killing development as before, we know that $U$ belongs to the
Killing horizon $\mathcal{N}_{\vec{\xi}}$.
Well-know properties of Killing horizons
imply $\nabla_{\mu}\lambda |_{\mathcal{N}_{\vec{\xi}}}
=2\kappa\xi_{\mu} |_{\mathcal{N}_{\vec{\xi}}}$, where $\kappa$
is the surface gravity and $\kappa^{2}=-2I_{1}$. Moreover,
(see e.g. theorem $7.3$  in
\cite{Heusler}) $\kappa$ is constant on each connected
component of the horizon in static spacetimes.
Therefore Lemma \ref{I1<0} implies that $I_1 <0$ on $U$.
Projecting the previous equation
onto $\Sigma$ it follows $D_{i}\lambda\big|_{V}=2\sqrt{-2I_{1}}Y_{i}\big|_{V}\neq 0$.
$\hfill \Box$

Fixed points are more difficult to analyze. We first need a Lemma
on the structure of $D_i N$ and $f_{ij}$ on  a fixed point.
\begin{lema}\label{lemmaFixed}
Let $\kid$ be a static KID and $p \in \partial \{ \lambda > 0 \}$ be a fixed point. Then
\begin{equation}
D_i N |_p\ne 0 \nonumber
\end{equation} 
and
\begin{equation}
\left. f_{ij} |_p = \frac{b}{Q} \left ( D_i N X_j - D_j N X_i \right )\right |_{p} \label{fijp}
\end{equation}
where $b$ is a constant, $X_i$ is unit and orthogonal to $D_i N |_p$ and $Q = \sqrt{D_i N D^i N}$.
\end{lema}
{\it Proof.} From (\ref{killingform}),
\begin{equation}\label{FF}
I_{1}=F_{\mu\nu}F^{\mu\nu}=f_{ij}f^{ij}-2
\left( D_{i}N+K_{ij}Y^{j} \right)\left( D^{i}N+K^{ik}Y_{k} \right).
\end{equation}
and $D_i N  |_p \neq 0$ follows directly from $I_1 |_p < 0$ (Lemma \ref{I1<0}).
For the second statement, let $u_i$ be unit and satisfy
$D_i N  = Q u_i$ in a suitable neighbourhood of $p$.
Consider (\ref{static5}) in the region $N \neq 0$, which
gives
\begin{eqnarray}
f_{ij} = -2 N^{-1} Y_{[i}\left( D_{j]}N+K_{j]k}Y^{k} \right). \label{fYN}
\end{eqnarray}
Since $|\vec{Y}|/N$ stays bounded in the region $\{ \lambda > 0 \}$,
it follows that the second summand tends to zero
at the fixed point $p$. Thus,
let $X_1^i$ and $X_2^i$ be any pair of vector fields orthogonal to
$u_i$. It follows by continuity that $f_{ij} X_1^i X_2^j |_p =0$.
Hence for any orthonormal basis $\{ u_i, X_i, Z_i \}$
at $p$ it follows $f_{ij} X^i Z^j |_p =0$ (because $\vec{X}$ and $\vec{Z}$ can be extended
to a neighbourhood of $p$ while remaining orthogonal to $\vec{u}$). Consequently
$f_{ij} |_p = (b/Q)  ( D_i N  X_j - D_j N  X_i )
+ ( c/Q)  ( D_i N  Z_j - D_j N  Z_i  ) |_{p}$
for some constants $b$ and $c$. A suitable rotation in the $\{X_i,Z_i \}$ plane
allows us to set $c=0$ and (\ref{fijp}) follows. $\hfill \Box$

Lemma \ref{I1<0} and expression (\ref{FF}) prove $D_i N D^i N |_p >
(1/2) f_{ij} f^{ij} |_p$,
or, by (\ref{fijp}), $Q^2|_p > b^2$. This will be used later.

An immediate consequence of this Lemma is that the set of fixed points, if open, is a smooth surface.
In fact, we will prove that this surface
is totally geodesic in $(\Sigma,g)$ and that the pull-back of the second fundamental form $K_{ij}$
vanishes there. This means from a spacetime perspective, i.e. when the initial data set is embedded
into a spacetime, that this open set of fixed points is  totally geodesic as a spacetime submanifold.
This is of course well-known in the spacetime setting from Boyer's results \cite{Boyer}, see
also \cite{Heusler}. In our initial data context, however, the result must be proven from scratch
as no Killing development is available at the fixed points.

\begin{proposition}
\label{Totally geodesic}
Let $\kid$ be a static KID
and assume that the set $\partial\{ \lambda>0 \}$ is
non-empty.
If $B\subset \partial\{\lambda>0 \}$ is open and consists of fixed points, then
$B$ is a smooth surface. Moreover,
the second fundamental form of $B$ in $(\Sigma,g)$ vanishes and $K_{AB} \big |_{B}=0$
\end{proposition}
{\it Proof.} On every point of $B$ we have $N =0$ and $D_i N \neq 0$, so $B$
is a smooth surface.

To prove the other statements,
let us introduce local coordinates $\{ u,x^{A} \}$
on $\Sigma$ adapted to $B$ so that $B\equiv\{ u=0 \}$ and let us prove that the linear
term in a Taylor expansion for $Y^i$ vanishes identically. Equivalently,
we want to show that $u^{j}D_{j}Y_{i} |_{B}=0$ for
$\vec{u} = \partial_u$  (recall that on $B$ we have
$Y_{i} |_{B}=0$ and this covariant derivative coincides with the partial derivative).
Note that $D_{i}Y_{j} |_{B}=f_{ij}$, so that
$u^{i} u^j D_{i}Y_{j} |_{B} =0 $ being the
contraction of a symmetric and an antisymmetric tensor. Moreover, for the tangential
vectors $e^i_A = \partial_A$ we find
$u^{j} e^i_A D_{i}Y_{j} |_{B} = u^j \partial_A Y_j =0$ because $Y_j$ vanishes all along $B$.
Consequently $u^{i}\partial_{i}Y_{j}|_{B}=0$. Hence, the Taylor expansion reads
\begin{eqnarray}\label{expansions}
N&=&G(x^{A})u+O(u^{2}),\nonumber \\
Y_{i}&=&O(u^{2}).
\end{eqnarray}
Moreover, $G\ne0$ everywhere on $B$  because substituting this Taylor expansion
in (\ref{I1}) and taking the limit $u \rightarrow 0$ gives
$I_{1} |_{B}=-2g^{uu}G(x^{A})^{2}$ and we know 
that $I_{1} |_{B} \neq 0$ from Lemma \ref{I1<0}.

We can now prove that $B$ is totally geodesic and that $K_{AB}=0$. For the first,
recall (\ref{DDN}). The Taylor expansion above gives $f_{ij}|_{B}=0$ and obviously 
$N$ and $\vec{Y}$ also vanish on $B$. Hence
$D_{i}D_{j}N |_{B}=0$.
Since, by Lemma \ref{lemmaFixed}, $D_{i}N |_{B}$ is proportional to the unit normal to $B$
and non-zero, then $D_{i}D_{j}N |_{B}=0$ is precisely the condition that
$B$ is totally geodesic. In order to prove $K_{AB}|_{B}=0$, we only need to
substitute the Taylor expansion (\ref{expansions}) in the $A,B$ components of
(\ref{kid1}). After dividing by $u$ and taking the limit $u \rightarrow 0$,
$K_{AB} |_{B}=0$ follows directly. $\hfill  \square$

When $\partial \{ \lambda >0 \}$ contains fixed points not lying on open sets, this boundary
is {\it not} a smooth surface in general.
Consider as an example
the Kruskal extension of the Schwarz\-schild
black hole and choose one of the asymptotic regions where the static Killing field is timelike. 
Its boundary consists of one half of the black hole event horizon, one half 
of the white hole event horizon and the bifurcation surface connecting both. 
Take an initial data set $\Sigma$ that intersects the bifurcation surface
transversally and consider the connected component
of the subset $\{ \lambda >0 \}$ within $\Sigma$ contained in the chosen asymptotic region. 
Its boundary is non-smooth because it has a 
corner on the bifurcation surface where the black hole event horizon and the white hole event 
horizon intersect (see example of Figure 1, where one spatial dimension has been suppressed).
\begin{figure}

\begin{center}
\includegraphics[width=9cm]{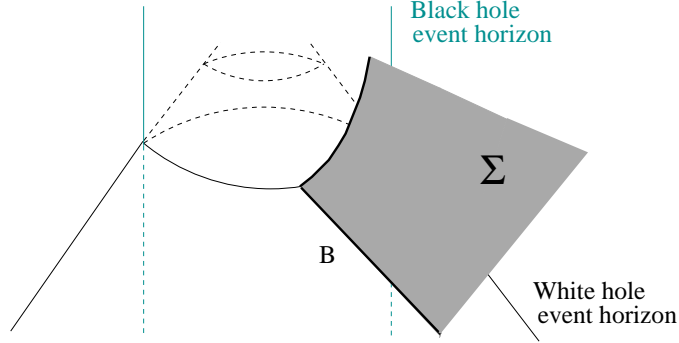}
\caption {An example of non-smooth boundary $B=\partial\{\lambda>0\}$ in an initial data set
$\Sigma$ of Kruskal spacetime with one dimension suppressed.
The region outside the cylinder and the cone corresponds to 
one asymptotic region of the Kruskal spacetime. The initial data set $\Sigma$ intersects
the bifurcation surface transversally. The shaded region corresponds to the intersection of $\Sigma$
with the asymptotic region, and is in fact a connected component of the subset $\{\lambda > 0 \} \subset
\Sigma$. Its boundary is non-smooth at the points lying on the bifurcation surface.}
\end{center}

\end{figure}
We must therefore add some condition
on $\partial \{ \lambda>0 \}$  in order to
guarantee that this boundary does not intersects both a black and a white hole event horizon.
In terms of the Killing vector, this requires that $\vec{Y}$ points only to one side
of $\partial\{\lambda>0\}$. Proposition \ref{maldito} suggests that the
condition we need to impose is
$Y^{i} D_{i} \lambda \big|_{\partial\{\lambda>0 \}}\ge0$
or $Y^{i} D_{i} \lambda \big|_{\partial\{\lambda>0 \}}\le0$.
This condition is in fact sufficient
to show that $\partial \{ \lambda > 0\}$ is a smooth surface. More precisely

\begin{proposition}\label{C1}
Let $\kid$ be a static KID
and consider a connected component $E$ of $\{ \lambda>0 \}$. If $Y^i D_i \lambda \geq 0$
or $Y^i D_i \lambda \leq 0$ on each connected component of $\E = \partial E$, then $\E$ is at least $C^1$.
\end{proposition}
{\it Proof.} Both cases are similar so only the case
$Y^i D_i \lambda  \geq 0$ will be proven. 
If there are no fixed points, the result
follows from Lemma \ref{surfaceNoFixed}. Let us therefore assume that
there is at least one fixed point $p$. 
The idea of the proof is to show that 
$Y^i D_i \lambda  \geq 0$ forces $b=0$ in (\ref{fijp}) from which smoothness will follow.
We argue by contradiction. Assume $b \neq 0$ in (\ref{fijp}). In a neighbourhood
of $p$, $D_i N \neq 0$ and we can use $x^1=N$ as a coordinate. Choosing coordinates $x^A = \{x,y \}$
on the
slice $\{ N=0 \}$  and extending them as constants along $D_i N$, the metric
$g$ takes the local form
\begin{eqnarray}
ds^2 = \frac{1}{Q^2} dN^2 + g_{AB} (N, x^C) dx^A dx^B. \label{metricatp}
\end{eqnarray}
Let us further choose $x^A$, so that $x^A(p)=0$, $g_{AB} |_p = \delta_{AB}$
and $dy |_p = \bm{X}$, where $\bm{X}$ is
the one-form appearing in (\ref{fijp}). Expanding $Y_i$ in Taylor series
we get $Y_i = s_i (x^A) + W_i(x^A) N + O(N^2)$ with $s_i (x^A=0)=0$ since $p$ is a fixed point.
Since $\frac{1}{N} (Y_i D_j N - Y_i D_j N)$ must have a finite limit at $p$ (it must in fact
coincide with $f_{ij} |_p$, see (\ref{fYN})),
it follows easily that $s_A =0$ on some neighbourhood of $p$. Restricting ourselves
to such neighbourhood, we have
$Y_i = \delta_i^1 r(x^A) + W_i (x^A) N + O(N^{2})$ for some function $r(x^A)$.
At $p$, we have $\partial_i Y_j |_p = D_i Y_j |_p = f_{ij} |_p$ because $D_{(i} Y_{j)}=0$
from (\ref{kid1}). Hence
\begin{eqnarray*}
\partial_i Y_j |_p = \left . \partial_A r \delta_i^A \delta_j^1 + W_j \delta_i^1
\right |_p = \frac{b}{Q_0}  \left (
\delta_i^1 \delta_j^{2} - \delta_j^1 \delta_i^{2} \right ),
\end{eqnarray*}
where (\ref{fijp}) has been used in the second equality and $Q_0 \equiv Q |_p$.
Hence $\partial_A r |_{p} = - \frac{b}{Q_0} \delta_A^{2}$,
$W_A |_p = \frac{b}{Q_0} \delta_A^{2}$ and $W_1 |_p =0$. Consequently
\begin{eqnarray*}
Y_1 = - \frac{b}{Q_0} y + O(2), \hspace{2cm} Y_A = \delta_A^{2} \frac{b}{Q_0} N + O(2)
\end{eqnarray*}
And then $\lambda = ( 1 - \frac{b^2}{Q_0^2} ) N^2
- b^2 y^2  + O(3)$. Recall that a connected
component of $\{ \lambda > 0\}$ has either $N >0$ or $N<0$ everywhere.
Let us choose $N>0$ for definiteness (the other case is
similar). The boundary of this region has $N \geq 0$. Moreover,
using $Q_0^2 > b^2$ and the expression above for $\lambda$,
it follows that, if $b \neq 0$, then $N$ vanishes
on the boundary of $\{ \lambda>0 \}$ 
only when  $y=0$.
A direct calculation using the metric (\ref{metricatp})
gives now $Y^i D_i \lambda = - 2 b Q_0 N y + O(2)$. Thus,
$Y^i D_i \lambda$ on the boundary changes sign with $y$ whenever $b \neq 0$. Consequently
the hypothesis of the proposition demands $b=0$ and therefore $f_{ij} |_p =0$.

It only remains to show that in these circumstances, $\E$ is $C^1$. We now have
$D_i D_j \lambda |_p = 2D_i N D_j N |_p$ and therefore $p$ is a degenerate critical point
for $\lambda$. The Gromoll-Meyer splitting Lemma \cite{Morse}
implies that there exists coordinates $\{v,x',y'\}$
in a neighborhood of $p$ such that $p = \{v=0,x'=0,y'=0\}$ and
$\lambda  = v^2 - q(x',y')$, for a smooth function $q$ satisfying $q(p) =0$, $D_i q |_p =0$
and $\mbox{Hess} (q) |_p =0$. But then, the boundary $\E$ is locally
defined by $v = \sqrt{q}$ (or with the minus sign,
depending on which connected  component is taken). The conditions on $q$ imply that $\E$ is $C^1$ at $p$.
$\hfill \Box$

Knowing that the surface is differentiable, our next aim is to show that, under suitable circumstances
it is in fact a MOTS. This is the content of our last proposition in this Section.

\begin{proposition}
\label{is_a_MOTS}
Let $\kid$ be a static KID
and consider a connected component $E$ of $\{ \lambda>0 \}$ with compact
boundary $\E = \partial E$. Assume
\begin{itemize}
\item[(i)] $N Y^i D_i \lambda|_{\E} \geq 0$  if $\E$ contains at least one fixed
point.
\item[(ii)]
$N Y^i m_i |_{\E} \geq 0$  if $\E$ contains no fixed point,
where $\vec{m}$ is the unit normal pointing towards $E$.
\end{itemize}
Then $\E$ is a MOTS with respect
to the direction pointing towards $E$.
\end{proposition}

{\em Remark.} If the inequalities in (i) and (ii) are reversed, then $\E$ is a MOTS with respect to
the unit normal pointing {\it outside} of $E$.

{\it Proof.}
Consider first the case when $\E$ has at least one fixed point.
Since $N$ is nowhere zero on $E$,
it must be either non-negative or non-positive
everywhere on $\partial E$. The hypothesis $N Y^i D_i \lambda |_{\E} \geq 0$
then implies either $Y^i D_i \lambda |_{\E} \geq 0$ or
$Y^i D_i \lambda |_{\E} \leq 0$ and Proposition \ref{C1} shows that $\E$ is a differentiable
surface. Let $\vec{m}$ be the unit normal pointing towards $E$ and $p$ the
corresponding mean curvature.  Being $\E$ also compact by hypothesis, it only remains to
show that $p + \gamma^{AB} K_{AB} =0$.

Let us start by proving that $\vec{Y}$ is everywhere orthogonal to
$\E$. At the fixed points, this is trivial as $\vec{Y} =0$.
For the open (possible empty) set $V$ of non-fixed points,
we can construct the Killing development of a suitable neighbourhood
and apply the Vishveshwara-Carter Lemma
\ref{carter} to show that $V$ lies on a Killing Horizon and therefore
 $\vec{\xi}^{T}\big|_{\E}= \vec{Y}^{T}\big|_{\E}=0$.
Moreover,
Killing horizons necessarily have
vanishing null expansion along to $\vec{\xi}$ and
equations   (\ref{killingdecomposition}) and (\ref{h}) imply
\begin{eqnarray}
\label{ExpansionKH}
\theta_{\vec{\xi}}= Y_{\mu}p^{\mu}  +N K_{AB}\gE^{AB}  |_{V} =0.
\end{eqnarray}
Now, Proposition \ref{maldito} implies that $D_i \lambda \neq 0$ on $V$
so that  $D_i \lambda |_{V} = H m_i |_{V}$, for a positive function $H$.
Using the fact that $\vec{Y}$ is parallel to $\vec{m}$, hypothesis (i)
implies $\vec{Y} |_{V} = N \vec{m} |_{V}$.
Dividing (\ref{ExpansionKH}) by $N$ it follows that $V$ has vanishing outer null
expansion with $\vec{m}$ pointing to the outer direction.
The same conclusion holds by continuity on isolated fixed points.
Open sets of fixed points
are immediately covered by Proposition \ref{Totally geodesic} because
this set is then totally geodesic and $K_{AB}=0$, so that both null expansions
vanish.

For the case that all points in $\E$ are non-fixed,
we know first of all that $\E$ is smooth from Lemma \ref{surfaceNoFixed}, and hence
$\vec{m}$ exists (this means in particular that hypothesis (ii) is well-defined).
The same argument as before shows that
$\vec{Y}$ is proportional to $\vec{m}$ and hypothesis (ii)
implies $\vec{Y} = N \vec{m}$ everywhere, so (\ref{ExpansionKH})
implies, as before, that the expansion along $\vec{m}$ vanishes.
$\hfill \Box$

\section{Main Results}
\label{mainresults}

In 2005 P. Miao \cite{M} proved a uniqueness theorem which generalized the usual uniqueness theorem for
static black holes by replacing the assumption of a black hole simply by the existence
of a minimal surface. More precisely, Miao worked with KIDs which are (i) time-symmetric
(which are defined by $K_{ij}=0,Y_i = 0$), (ii) vacuum
and (iii) asymptotically flat. The latter is defined as follows
(recall that
a function is said to be $O^{(k)}(r^{n})$, $k\in\mathbb{N}$ if $f(x^{i})=O(r^{n})$, $\partial_{j}f(x^{x^{i}})=O(r^{n-1})$ and so on for all derivatives up to an including the k-th ones).

\begin{defi}
\label{asymptoticallyflat1}
A KID $\kid$ is {\bf asymptotically flat} if
$\Sigma=K \cup \Sigma^{\infty}$, where $K$ is a compact set and
$\Sigma^{\infty}=\underset{a}{\bigcup}\Sigma^{\infty}_{a}$ is a finite union
with each $\Sigma^{\infty}_{a}$,
called an ``asymptotic end'' being diffeomorphic
to $\mathbb{R}^3 \setminus
\overline{B_{R_{a}}}$, where $B_{R_{a}}$ is an open ball of radius $R_{a}$. Moreover, in
the Cartesian coordinates
induced by the diffeomorphism, the following decay holds
\begin{eqnarray}
    N-A_{a} = O^{(2)}(1/r),\qquad g_{ij}-\delta_{ij}&=&O^{(2)}(1/r),\nonumber\\
    Y^{i}_{a}-C^{i}_{a}=O^{(2)}(1/r), \qquad\qquad
    K_{ij}&=&O^{(2)}(1/r^{2}).\nonumber
\end{eqnarray}
where $A_{a}$ and $\{C^{i}_{a} \}_{i=1,2,3}$
are constants such that $A^2_{a}-\delta_{ij} C^{i}_{a} C^{j}_{a}>0$
for each $a$, and $r=\left(x^{i}x^{j}\delta_{ij} \right)^{1/2}$.
\end{defi}

The condition on the constants $A_a, C^i_a$ is imposed to ensure that
the KID is timelike near infinity on each asymptotic end.

Miao's theorem reads
\begin{thr}[Miao,  2005 \cite{M}]\label{miao}
Let $(\Sigma,h,K=0;N,\vec{Y}=0,\tau)$ be a time-symmetric, vacuum and
asymptotically flat KID with a compact minimal surface
(i.e. a surface of vanishing mean curvature)
which bounds an open domain $W\subset\Sigma$.

Then $(\Sigma\setminus W,h)$ is isometric to
$\left(\mathbb{R}\setminus
 B_{m/2}(0),g_{Schw}=\left( 1+\frac{m}{2|x|}\right)^4\delta_{ij}\right)$ for some $m >0$.
\end{thr}

{\em Remark 1.} The metric $g_{Schw}$ is the induced metric
on the $\{t=0\}$ slice of the Schwarzschild
spacetime with mass $m$, outside and including the horizon.

{\em Remark 2.} Actually, Miao's theorem \cite{M} deals with KIDs which
have minimal boundaries. The formulation given above is
in principle weaker but more suitable for our purposes.

One way of understanding the contents of this theorem is that a time-symmetric, vacuum and
 asymptotically flat Killing initial data set which contains a bounding minimal
surface must in fact be a black hole and
that the bounding minimal surface cannot penetrate into the exterior region defined
as the connected component of $\{ \lambda>0 \}$ containing infinity. Thus, the minimal surface will be hidden
inside the black hole and the usual uniqueness theorem for black holes implies that the slice
must be Schwarzschild.
The aim of the theorems below is to extend this result showing that bounding MOTSs cannot penetrate into the ``exterior''
region where the Killing vector is timelike. 
Our proof is strongly based on a powerful theorem recently proved by L. Andersson and
J. Metzger \cite{AM}, extending previous work by Schoen \cite{Schoen}. Let us
first state this theorem.

\subsection{Andersson \& Metzger Theorem}\label{subsectionAM}

The key object for the Andersson \& Metzger theorem is
that of {\it weakly outer trapped region}, defined as follows.

\begin{defi} Let $\id$ be an initial data set with an untrapped barrier surface
$\Sb=\partial \Db$. An open set $\Omega\subset\Db$ is called
{\bf weakly outer trapped set} if $\partial \Omega$ is a smooth embedded closed surface
that is weakly outer trapped w.r.t. the normal pointing outside  $\Omega$.
\end{defi}
\begin{defi} Let $\id$ be an initial data set.
The {\bf weakly outer trapped region}, $T$, is defined as the union of all the
weakly outer trapped sets.
\end{defi}
\begin{thr}[Andersson \& Metzger \cite{AM}] \label{AM}
Let $\id$ be a smooth, initial data set containing an untrapped barrier surface
$\Sb=\partial\Db$ with $\overline{\Db}$ complete. Let $T$ be the weakly outer trapped region.
Then either $T=\emptyset$ or
$\partial T$ is a smooth stable MOTS.
\end{thr}

{\em Remark.} The definition of stable MOTS can be found in
\cite{AMS}. For the purposes of this paper, we only need the property that
$\partial T$ is a MOTS.

\subsection{Main Theorems}\label{mainth}

The idea of the proof of the theorems below
is to assume the existence of a bounding MOTS $\mots$ in the exterior region,
and use Andersson-Metzger theorem to pass
to the outermost MOTS given by $\partial T$, which by construction must be on or
outside $\mots$.
Then, using stationarity and the null energy condition (NEC)
we construct another MOTS strictly outside $\partial T$ therefore getting
a contradiction.

Let us first recall the definition of null energy condition (NEC).
\begin{defi}\label{NEC}
A Killing initial data set $\kid$ satisfies the null energy condition (NEC)
if for all $p\in \Sigma$ the tensor $G_{\mu\nu} \equiv \rho n_{\mu} n_{\nu} + J_{\mu} n_{\nu}
+ n_{\mu} J_{\nu} + \tau_{\mu\nu} |_p $ on $T_{p}\Sigma\times\mathbb{R}$
satisfies that $G_{\mu\nu} k^{\mu}k^{\nu}\ge 0$ for any null vector
$\vec{k} \in T_{p}\Sigma\times\mathbb{R}$.
\end{defi}

The first of our main theorems involves KIDs having an untrapped barrier surface with
no further restriction.
Staticity is not required for this result.
\begin{thr}\label{theorem1}
Let $\kid$ be a KID containing an untrapped barrier $\Sb=\partial \Db$ and satisfying NEC. Assume
that $\overline{\Db}$ is complete.

Then, there exists
{\bf no bounding MOTS} $\mots = \partial \D$ with $\D \subset\Db$
satisfying
$\Sigma \setminus \D \subset \{ \lambda \geq 0 \}$
provided
$\mots \cap \{ \lambda > 0 \} \neq \emptyset$.
\end{thr}
\begin{figure}[h]

\begin{center}
\includegraphics[width=9cm]{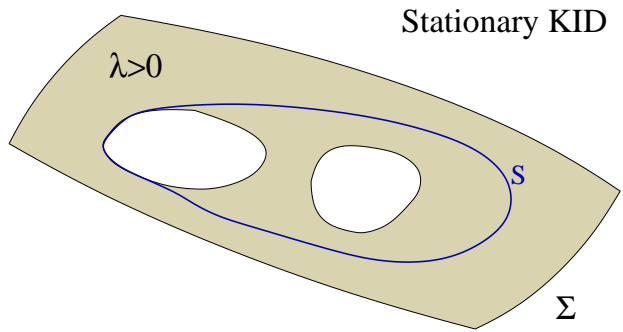}
\caption {Theorem \ref{theorem1} excludes the possibility that a MOTS like the one in
the figure exists. The shaded area corresponds to the region where $\lambda$ is positive.}
\end{center}

\end{figure}

{\em Remark 1.} In short, the conditions of the theorem demands that
the bounding MOTS is such that that Killing vector is causal everywhere
on its exterior and timelike at least somewhere. When MOTS is replaced by the stronger condition
of being a marginally trapped surface with one of the expansions non-zero somewhere, then
this theorem can be proven by a simple argument based on the first variation of area \cite{MS}.
In that case, the assumption of $S$ being bounding becomes unnecessary. It would be
interesting to know if theorem \ref{theorem1} holds for arbitrary MOTSs, not necessarily bounding.

{\em Remark 2.} When the KID is asymptotically flat, the surface at constant $r$
on each one of the asymptotic ends is, for large enough $r$, outer untrapped. Thus,
the domain $\Db$ can be taken as large as desired and in particular so that
it contains any given MOTS $\mots$. In that case, the theorem asserts
that there exists {\bf no bounding MOTS} $\mots = \partial \D$ in $\Sigma$ such that
$\Sigma \setminus \D \subset \{ \lambda \geq 0 \}$ and
$\mots \cap \{ \lambda > 0 \} \neq \emptyset$.

Two immediate, but useful corollaries follow.

\begin{corollary}\label{corollary}
Assume that $\{ \lambda>0 \}=\Sigma$ on an asymptotically flat KID $\kid$
satisfying NEC, then
there exists no bounding MOTS in $\Sigma$.
\end{corollary}

\begin{corollary}\label{corollary2}
Let $\id$ be an initial data set for Minkowski space, then there exists no bounding MOTS in $\Sigma$.
\end{corollary}

It is obvious that the second Corollary is a particular case
of the first one because the $\partial_t$
in Minkowskian coordinates is strictly stationary everywhere, in particular on $\Sigma$.
The non-existence result of a bounding MOTS in a Cauchy surface
of Minkowski spacetime is however, well-known
as this spacetime is obviously regular predictable (see \cite{HE} for definition) and then the
proof of Proposition $9.2.8$ in \cite{HE} gives the result.

{\it Proof.} Under the conditions of the theorem, it is clear that $\D$ belongs to the
weakly outer trapped region $T$. Hence $\partial T \subset
\{\lambda  \geq 0 \}$ with at least one point having $\lambda > 0$.

Assume first that no point in $\partial T$ is fixed. Then $N \neq 0$ in some neighbourhood of
$\partial T$ and we can construct the Killing development there. Since
the KID satisfies NEC, so does the Killing development (the Einstein tensor is Lie constant
along $\partial_t$). We can now consider the action of the local isometry
group generated by the Killing vector $\vec{\xi}$, which is causal on $\partial T$.
Let $\gamma$ be the group parameter and  drag $\partial T$ with the local isometry
a constant negative amount $\gamma_0$ of the group parameter
($\gamma_0$ can be chosen small enough so that the local group exists up to
this value). Denote the image surface by $\mots'$.
Since $\vec{\xi}$ is future directed, $\mots'$
lies strictly in the causal past of $\partial T$. Since geometric properties remain
unchanged under the action of an isometry, $\mots'$ is still a MOTS in the spacetime.
Let $\vec{l}$ be the Lie dragging of $\vec{n} + \vec{m}$ onto $\mots'$ and
consider the null geodesics starting
on $\mots'$ and with tangent vector $\vec{l}$. This generates a null hypersurface which is
smooth near enough $\mots'$. Null hypersurfaces $\N$ ruled by a null vector $\vec{l}$
are endowed with a null expansion $\theta$
which has the property that any spacelike surface contained in $\N$ has null expansion with respect
to $\vec{l}$ equal to $\theta$ (see e.g. \cite{Galloway}). Moreover $\theta$ obeys the
Raychaudhuri equation: let $\beta$ be the affine parameter associated to $\vec{l}$ such that
$\beta=0$ on $\mots'$. Then
\begin{eqnarray*}
\frac{d \theta }{d \beta} = - \frac{1}{2} \theta^2 - \sigma^2 - G_{\mu\nu} l^{\mu} l^{\nu},
\end{eqnarray*}
where $\sigma$ is the shear scalar of $\N$. Using NEC, all terms
in the right hand side are non-positive. Since $\theta=0$ at $\beta=0$,
it follows that $\theta$ is non-positive
in the future of $\mots'$ in $\N$. In general $\N$ will develop singularities in the future,
however, the first singularity will occur for a finite value of $\beta$, which is independent of
$\gamma_0$ (i.e. the amount we shifted $\mots$ to the past). It is therefore clear that
by choosing $\gamma_0$ small enough,
$\mots'' \equiv \Sigma \cap \N$ will be a smooth surface (obviously lying in the
future of $\mots'$). By Raychaudhuri, this surface has non-positive outer expansion.
Moreover, by construction $\mots''$ is
also bounding for small enough $\gamma$ (because it is constructed by a continuous
deformation of $\partial T$, which is bounding) and lies strictly in the
exterior of $\partial T$ (because on at least one point of $\mots$
$\vec{\xi}$  is timelike). But this gives a contradiction since $\mots''$ must belong to $T$ by definition of $T$.

When $\partial T$ has fixed points, we cannot guarantee the existence of the Killing development
on those points.
However, letting $U \subset \partial T$ be the set of non-fixed points
(which is obviously non-empty and open within $\partial T$), such development still exists
in a neighbourhood of $U$. In this portion we can repeat the construction above and define
$\mots'' \subset \Sigma$.
$N$ and $\vec{Y}$ being smooth and approaching zero at the fixed points,
it follows easily that $\mots''$ and the set of fixed points
$\partial T \setminus U$ will join smoothly and will therefore define a surface, which we still
denote by $\mots''$. Moreover, this
is still a bounding weakly outer trapped (it is still a continuous deformation of $\mots$)
and therefore must be contained in $T$. But
at least one point in $\partial T$ lies in $\{\lambda >0 \}$, so that $U$ must be  non-empty
and $\mots''$ has at least one portion strictly outside $\partial T$, which gives the desired
contradiction.
$\hfill \square$

Notice that Corollary \ref{corollary} together with Proposition \ref{Totally
geodesic} already imply the uniqueness part of Miao's theorem. Indeed,
Corollary
\ref{corollary} asserts that the existence of a bounding minimal
surface in an asymptotically flat KID implies that $\partial\{
\lambda>0 \}$ is non-empty and compact, while Proposition \ref{Totally
geodesic} states that the set $\partial\{ \lambda>0 \}$ in such a
time-symmetric KID is a totally geodesic surface in $\Sigma$ (in the
time-symmetric case this was already known, see e.g. \cite{Corvino}).
Thus, the usual uniqueness theorem for
vacuum black holes implies that the exterior of this totally geodesic
surface coincides with the $r > 2m$ region of the $\{t=0 \}$ slice in
Schwarzschild coordinates. However, Miao's result also states that
the original minimal surface cannot
penetrate into the exterior region $\{\lambda>0\}$. This last part
is recovered (and extended) in our next theorem, where
we show that bounding MOTSs cannot penetrate into the exterior
$\{ \lambda > 0 \}$ region in a static KID provided a suitable untrapped barrier
surface exists (in particular in the asymptotically flat case).

\begin{thr}\label{theorem2}
Let $\kid$ be a static KID containing an untrapped barrier surface $\Sb=\partial\Db$
and satisfying
NEC.
Assume that $\overline{\Db}$ is complete and $\Sigma \setminus \Db \subset \{ \lambda > 0 \}$. Let $\U$ be the
connected
component of $\{ \lambda>0  \}$ containing $\Sigma \setminus \Db$.
Suppose that $\partial\U$ is closed and
\begin{itemize}
\item[(i)] $N Y^i D_i \lambda|_{\partial \U} \geq 0$  if $\partial \U$ contains at least one fixed
point.
\item[(ii)] $N Y^i m_i |_{\partial \U} \geq 0$  if $\partial \U$ contains no fixed point,
where $\vec{m}$ is the unit normal pointing towards $\U$.
\end{itemize}

Then, there exists {\bf no bounding MOTS} $\mots = \partial \D$, with $\D \subset \Db$, intersecting $U$.
\end{thr}

\begin{figure}[h]

\begin{center}
\includegraphics[width=9cm]{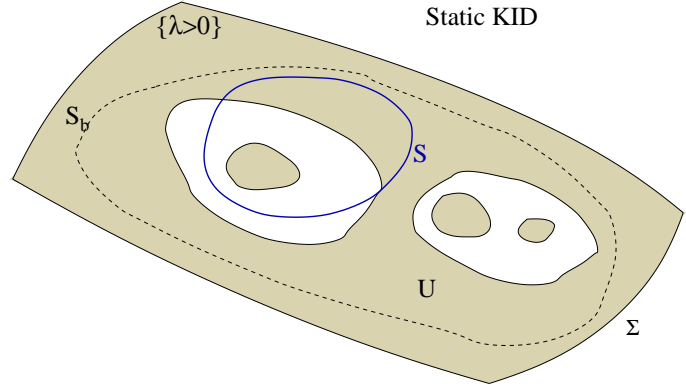}
\caption {Theorem \ref{theorem2} forbids the existence of a bounding MOTS S like the one in the
figure. The shaded area corresponds to the region where $\lambda$ is positive. $S_b$ is the
untrapped barrier and the region it encloses is $\Db$.}
\end{center}

\end{figure}

{\em Remark.} In asymptotically flat KIDs
the hypotheses of the theorem regarding the untrapped barrier and the compactness of
$\partial \U$ are automatically satisfied. Consequently, in this case
no bounding MOTS intersecting $U$ can exist, provided
(i) or (ii) hold.

{\it Proof.} First of all, we know from Proposition \ref{is_a_MOTS} that $\partial \U$ is a MOTS.
Assume there exists a bounding MOTS $\mots$ intersecting $\U$.
We know by
definition of $T$ that both $\partial \U$ and $\mots$ are contained in $T$. Therefore
$\Sigma \setminus T \subset \overline{\U}$ from which it follows that $\partial T \subset
\overline{\U}$ with at least one point in $U$. But then the same construction as in
Theorem \ref{theorem1} gives a contradiction. $ \hfill \square $

Clearly, this theorem recovers Miao's result
in the particular case of asymptotically flat time-symmetric vacuum KIDs containing a bounding minimal surface.
Notice that when all points in $\partial\U$ are fixed points
$Y^{i}D_{i}\lambda\big|_{\partial\U}$ is identically zero.

We finish this work remarking that the same effort used to 
prove the previous results leads to the following
Corollary.
\begin{corollary}
Theorems \ref{theorem1} and \ref{theorem2}, and Corollaries \ref{corollary} and \ref{corollary2} also hold
if ``bounding MOTS'' is replaced by {\bf bounding weakly outer trapped surface}.
\end{corollary}
{\it Proof.}
A bounding weakly outer trapped surface $S$ is included in the weakly outer trapped region $T$, so the same proof
as before applies. $ \hfill \square $

\section*{Acknowledgments}

We are very grateful to Lars Andersson and Jan Metzger for
letting us know their results on smoothness of the boundary of the trapped
region prior to publication. We also thank Lars Andersson and Walter Simon
for useful discussions and Jos\'e M. M. Senovilla for comments on the manuscript. Financial support under the projects
FIS2006-05319 of the Spanish MEC and
SA010CO5 of the Junta de Castilla y Le\'on is acknowledged. AC 
acknowledgments a Ph.D. grant (AP2005-1195) from the Spanish MEC.

\end{document}